\documentclass[aps,prd,amssymb,twocolumn,10pt]{revtex4-1}
\newcommand*\widebar[1]{%
  \hbox{%
    \vbox{%
      \hrule height 0.5pt 
      \kern0.5ex
      \hbox{%
        \kern-0.1em
        \ensuremath{#1}%
        \kern-0.1em
      }%
    }%
  }%
}
\usepackage{graphicx}
\usepackage{dcolumn}%
\usepackage{bm}%

\usepackage[version=3]{mhchem}
\usepackage{slashed}

\usepackage[bookmarks=false]{hyperref}

\begin{document}


\title{Contribution of $\sigma$ meson exchange to elastic lepton-proton scattering}


\author{Oleksandr Koshchii}
\email[]{koshchii@gwmail.gwu.edu}
\affiliation{The George Washington University, Washington, DC 20052, USA}

\author{Andrei Afanasev}
\email[]{afanas@gwu.edu}
\affiliation{The George Washington University, Washington, DC 20052, USA}


\date{\today}

\begin{abstract}
Lepton mass effects play a decisive role in the description of elastic lepton-proton scattering when the beam's energy is comparable to the mass of the lepton. The future Muon Scattering Experiment (MUSE), which is devised to solve the ``Proton Radius Puzzle'', is going to cover the corresponding kinematic region for a scattering of muons by a proton target. We anticipate that helicity-flip meson exchanges will make a difference in the comparison of elastic electron-proton vs muon-proton scattering in MUSE. In this article, we estimate the $\sigma$ meson exchange contribution in the $t$ channel. This contribution, mediated by two-photon coupling of  $\sigma$, is calculated to be at most $\sim 0.1 \%$ for muons in the kinematics of MUSE, and it appears to be about 3 orders of magnitude larger than for electrons because of the lepton-mass difference.
\end{abstract}

\maketitle

\section{Introduction}
Elastic lepton-nucleon scattering has proven to be a valuable tool to gain insight into the structure of the nucleon. Over the past few decades, development of experimental technologies has made it possible to reveal effects in observed cross sections or polarization asymmetries that are on the order of few tenths of a percent. Such precision of experimental measurements requires theoretical calculations of electromagnetic corrections to be done beyond the leading-order Born approximation. In particular, two-photon exchange (TPE) effects, which are found to contribute at the level of few percent \cite{PhysRevD.72.013008, PhysRevLett.91.142304}, appear to be important in this context. At momentum transfers $Q^2 \lesssim 1$GeV$^2$, TPE calculated within a hadronic framework that only includes nucleon-size effects \cite{PhysRevLett.91.142304} is in good agreement with new experimental data on charge asymmetries of electron vs positron scattering on the proton target \cite{PhysRevLett.114.062003, PhysRevLett.114.062005}. However, when we deal with a kinematic region where the lepton's mass $m$ cannot be neglected, the effects due to lepton helicity flip need to be taken into consideration, potentially leading to larger theoretical uncertainties that are not constrained by electron vs positron comparison. It should be noted that TPE calculations require the knowledge of a virtual Compton scattering amplitude on the nucleon, the Born terms of which were included in calculations of Ref. \cite{PhysRevD.90.013006}. Our objective is to consider inelastic terms that are most sensitive to the lepton mass, namely, $t$-channel exchange of scalar mesons.

The real part of the TPE amplitude, which affects the analyses of cross sections, can be studied through the difference between elastic lepton-nucleon and antilepton-nucleon scattering. Such a technique is going to be implemented, for instance, in the future Muon Scattering Experiment (MUSE) at Paul Scherrer Institute, Switzerland. MUSE is motivated by the recent measurements  \cite{pohl2010size, Antognini417} of the charge radius of a proton $r_{ch}$. Both results, $r_{ch} = 0.84184(67)$ fm and $r_{ch} = 0.84087(39)$ fm, respectively, were obtained using Lamb shift measurements in the muonic hydrogen atom and are inconsistent with older values of the ``radius'' collected in nonmuonic experiments. These nonmuonic measurements include two independent determinations of the charge radius of the proton: from elastic $e - p$ scattering and from Lamb shift measurements in the hydrogen atom. The combined electron-based result is $r_{ch} = 0.8775(51)$ fm \cite{RevModPhys.84.1527}. This means that the discrepancy between muonic and electronic results is $> 7 \sigma$. Because of this, the problem has been named the ``Proton Radius Puzzle'' and has led to the proposal \cite{gilman2013studying} for the MUSE experiment. This experiment will measure simultaneously elastic $\mu^\pm - p$ and $e^\pm - p$ scattering. This will enable experimentalists to compare $e$ vs $\mu$ charge radii measured in the same setting, as well as test several possible explanations of the puzzle, such as the lepton universality violation, physics beyond the Standard Model, and enhanced TPE contributions.

The goal of MUSE is to extract the charge radius of the proton with error bars similar to previous $e - p$ measurements. As a result, in order to have systematic uncertainties under control, the relative unpolarized cross section observables have to be calculated at a level of few tenths of per cent, making theoretical estimations of QED corrections, and TPE in particular, extremely important. These calculations are complicated by the fact that kinematic conditions of MUSE are such that the beam momenta in the $100-200$ MeV range do not allow us to use the ultrarelativistic (UR) limit ($m \rightarrow 0$) in muon estimations. Therefore, as it was pointed out above, the helicity-flip meson-exchange amplitudes will play an important role in TPE calculations. Precision requirements of MUSE suggest that the interference of these amplitudes with the one-photon exchange (OPE) amplitude can be substantial enough to be taken into account. We would expect the largest contribution there to be coming from the long-range light meson exchanges in the $t$ channel, making neutral pions look like the most tempting candidates for the leading contribution (see Fig. \ref{fig:1}). It appears, however, that the interference between the vector current, which represents the one-photon-exchange process, and the pseudoscalar current, which represents the pion exchange, is zero for the case of unpolarized particles. Consequently, we predict the largest nonvanishing $t$-channel interference contribution to be obtained from the scalar $\sigma$ [also known as $f_0(500)$] meson exchange. This exchange manifests itself in the nucleon's polarizabilities estimations \cite{ANDP:ANDP201400077} and in the D term of the nucleon's generalized parton distributions \cite{PhysRevD.60.114017}; for the detailed discussion of $f_0(500)$ properties, please see the review in Ref. \cite{Agashe:2014kda}. Besides that, the idea to consider scalar, pseudoscalar, and tensor meson exchanges in an elastic lepton-proton scattering was also discussed in Ref. \cite{PhysRevC.90.045205}. However, the authors of that paper considered only the scattering of ultrarelativistic electrons thereby neglecting the mass-dependent contribution.

In this paper, we estimate the leading $\sigma$ meson exchange contribution to the unpolarized scattering cross section in MUSE kinematics and show that the analogous (calculated to the same order in the fine structure constant) single-pion exchange contribution is zero.

\section{Elastic Lepton-Proton Scattering Formalism}

To describe the following elastic lepton scattering off of a proton
\begin{equation}\label{1.1}
    l(k_1) + p(p_1) \rightarrow l(k_2) + p(p_2)
\end{equation}
we will use the Mandelstam variables
\begin{equation}\label{1.2}
    s = (k_1 + p_1)^2, \ t = q^2 = (k_1 - k_2)^2, u = (k_1 - p_2)^2.
\end{equation}

\begin{figure}
    \includegraphics[scale=0.34]{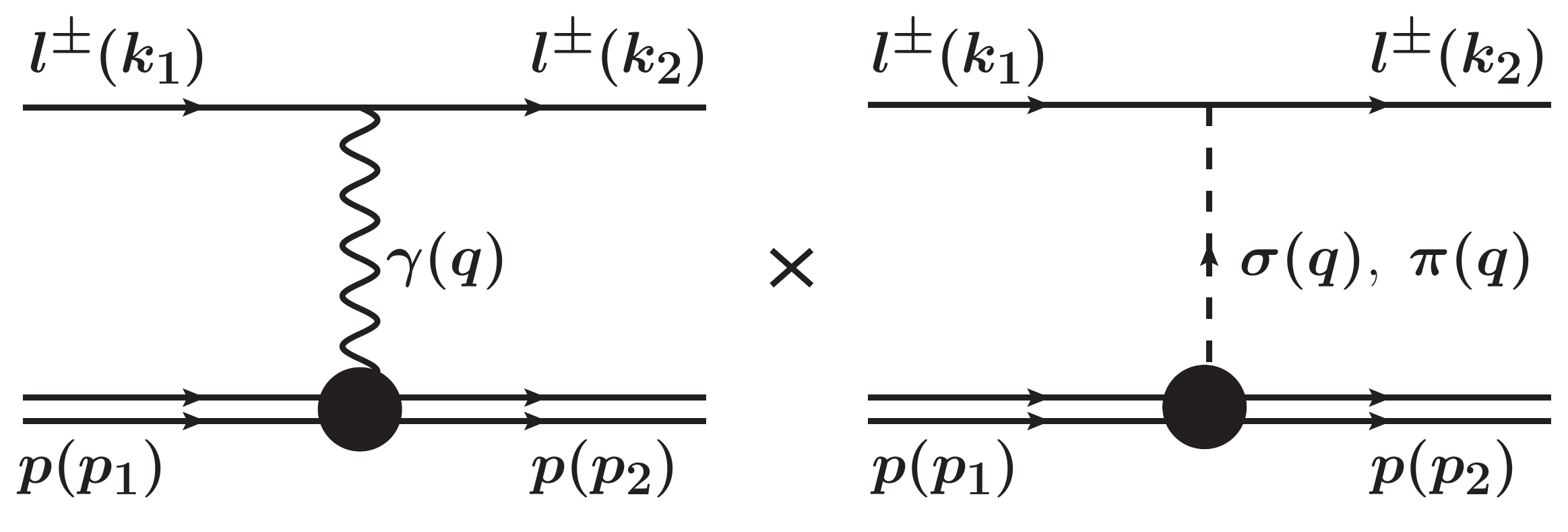}
    \caption{\label{fig:1}One-photon and one $\sigma$ ($\pi$) meson exchange diagrams}
\end{figure}

The left diagram in Fig. \ref{fig:1} represents the leading-order OPE contribution. The corresponding lepton and proton vector currents are given by
\begin{equation}\label{1.3}
    j^v_{\mu} = \bar{u} (k_2) \gamma_{\mu} u (k_1),
\end{equation}
\begin{equation}\label{1.4}
    J^v_{\mu} = \bar{U} (p_2) \Big( \gamma_{\mu} F_1(Q^2) + \frac{i \sigma_{\mu \nu} q_\nu}{2 M} F_2(Q^2) \Big)
    U (p_1),
\end{equation}
where $M$ is the mass of the proton, $F_1(Q^2)$ and $F_2(Q^2)$ are the Dirac and Pauli form factors, $\sigma_{\mu \nu} \equiv \frac {i}{2} [\gamma_\mu, \gamma_\nu]$, $Q^2 \equiv - q^2 > 0$.

The scalar (pseudoscalar) $\sigma$ ($\pi$) meson exchange process is described by the right diagram in Fig. \ref{fig:1}. Associated scalar (pseudoscalar) currents can be written as
\begin{equation}\label{1.5}
\begin{split}
    j^s = & f_s \bar{u} (k_2) u (k_1), \ \ \ \ \ \ \ \ j^p = f_p \bar{u} (k_2) \gamma_5 u (k_1), \\
    J^s = & g_s \bar{U} (p_2) U (p_1), \ \ \ \ \ \ J^p = g_p \bar{U} (p_2) \gamma_5 U (p_1),
\end{split}
\end{equation}
where $f_{s (p)} = f_{s (p)} (Q^2)$ and $g_{s (p)} = g_{s (p)} (Q^2)$  are the form factors that describe the coupling of $\sigma$ ($\pi$) to the lepton and proton, correspondingly. There is no information available about the $f_{s (p)}$ form factor. Therefore, we will introduce a theoretical model to estimate the leading contribution to this quantity. This model is discussed in Sec. \ref{1.100}.

The square of the matrix element that includes one photon, one scalar meson (mass $m_s$), and one pseudoscalar meson (mass $m_p$) exchanges is given by
\begin{equation}\label{1.7}
\begin{split}
    |\mathcal{M}|^2 = & |\mathcal{M}^v + \mathcal{M}^{s} + \mathcal{M}^{p}|^2 \\
    \approx & |\mathcal{M}|_{1 \gamma}^2 + 2 \ \mathrm{Re} [\mathcal{M}^v {\mathcal{M}^s}^*] + 2 \ \mathrm{Re} [\mathcal{M}^v {\mathcal{M}^p}^*],
\end{split}
\end{equation}
where for the $l^\pm$ scattering we have
\begin{equation*}\label{1.7'}
\begin{split}
   & \mathcal{M}_{1 \gamma} \equiv \mathcal{M}^v = \mp \frac{i e^2}{Q^2} \ j^v_{\mu} J^v_{\mu}, \\
   & \mathcal{M}^{s(p)} = - \frac{i}{Q^2 + m^2_{s(p)}} \ j^{s(p)} J^{s(p)}. \\
\end{split}
\end{equation*}
Note that in Eq. (\ref{1.7}) we neglected $|\mathcal{M}^s|^2$, $|\mathcal{M}^p|^2$, $\mathcal{M}^s {\mathcal{M}^p}^*$, and $\mathcal{M}^p {\mathcal{M}^s}^*$ terms. These terms are irrelevant for estimations within the required level of accuracy. The detailed derivation of the explicit form of the second term in Eq. (\ref{1.7}) as well as the proof that the third term in this equation is zero are given in Appendix \ref{1.300}.

\section{Born Approximation}

The most convenient frame to calculate the first term in Eq. (\ref{1.7}) is the Breit frame, our notations for which are summarized in the left column of Eq. (\ref{1.8}). The Breit frame is defined as the frame in which there is no energy transfer between the lepton and the proton $q = (0, \vec{q}_{{B}})$. It makes the square of the 4-momentum transfer to be simply related to the square of its spatial component: $q^2 = - {\vec{q}}^{\ 2}_{{B}}$. Consequently, the transition from the Breit frame to any other frame can be performed in a straightforward manner.
\begin{equation}\label{1.8}
    \overset{\mathsf{Breit \ Frame:}}{
    \begin{cases}
        p_1 = \big( E_{{B}}, - \frac {\vec{q}_{{B}}}{2} \big), \\
        p_2 = \big( E_{{B}}, \frac {\vec{q}_{{B}}}{2} \big), \\
        k_1 = (\varepsilon_{{B}}, \vec{k}_{1{B}}), \\
        k_2 = (\varepsilon_{{B}}, \vec{k}_{2{B}}).
    \end{cases}} \ \ \ \ \ \
    \overset{\mathsf{Lab \ Frame:}}{
    \begin{cases}
        p_1 = \big( M, 0 \big), \\
        p_2 = \big( E_2, \vec{p}_2 \big), \\
        k_1 = \big( \varepsilon_1, \vec{k}_{1} \big), \\
        k_2 = \big( \varepsilon_2, \vec{k}_{2} \big).
    \end{cases}}
\end{equation}

The details of deriving the result, using the Born approximation in the Breit frame, can be found in \cite{rekalo2002polarization}. The only peculiarity that needs to be taken into account due to the nonzero mass of the lepton is the modified energy of the lepton
\begin{equation*}
    \varepsilon_{{B}} = \sqrt{m^2 + \frac {Q^2}{4 \sin^2 \frac {\theta_{{B}}}{2}}},
\end{equation*}
where $\theta_{{B}}$ is the scattering angle in the Breit frame and $m$ is the lepton's mass.

To be consistent with MUSE, we assume that the beam and the target have no polarization preference. This means that the matrix element has to be summed over the final polarizations and averaged over initial ones. Then, the corresponding Breit frame result obtains the form
\begin{multline} \label{1.9}
    |\widebar{\mathcal{M}}|_{1 \gamma, B}^2 = \frac{4 M^2 e^4}{Q^2} \Big[ \frac{4 m^2}{Q^2} G_E^2(Q^2) + 2 \tau G_M^2(Q^2) \\
    + \Big( G_E^2(Q^2) + \tau G_M^2(Q^2) \Big) \cot^2 \frac{\theta_{{B}}}{2} \Big]
\end{multline}
with $\tau \equiv \frac{Q^2}{4 M^2} > 0$. The electric $G_E (Q^2)$ and magnetic $G_M (Q^2)$ Sachs form factors are defined by
\begin{equation}\label{1.110}
    G_E (Q^2) = F_1(Q^2) - \tau F_2(Q^2),
\end{equation}
\begin{equation}\label{1.111}
    G_M (Q^2) = F_1(Q^2) + F_2(Q^2).
\end{equation}

These form factors fall off similarly with $Q^2$ in the kinematic region of our interest ($Q^2 \lesssim 0.1$ GeV$^2$). Thus, to a good approximation, the underlying parametrization can be chosen to describe their $Q^2$ behavior \cite{povh1995particles},
\begin{equation*}
    G_E (Q^2) = \Big( 1 + \frac {Q^2}{\Lambda^2} \Big)^{-2}, \ \ \ G_M (Q^2) = \mu G_E(Q^2)
\end{equation*}
with $\Lambda^2 = 0.71 \ \mathrm{GeV}^2$ and the magnetic moment of the proton $\mu = 2.793$.

The Breit frame result Eq. (\ref{1.9}) can be converted to the Lab frame through the simple relation between the corresponding scattering angles; our notation for the Lab frame is shown in the right column of Eq. (\ref{1.8}). As a consequence, the Born approximation matrix element in the Lab frame is
\begin{multline}\label{1.10}
    |\widebar{\mathcal{M}}|_{1 \gamma}^2 = \frac{4 M^2 e^4}{Q^2} \Big[ \frac{4 m^2}{Q^2} G_E^2(Q^2) + 2 \tau G_M^2(Q^2) \\
    + \Big( G_E^2(Q^2) + \tau G_M^2(Q^2) \Big) \frac{4 \vec{k}_1^2 \ \vec{k}_2^2 }{Q^4 (1 + \tau)} \sin^2 \theta \Big],
\end{multline}
where $\theta$ is the scattering angle. From now on, all the expressions will be given in the Lab frame.

The differential cross section for the described process, Eq. (\ref{1.1}), can be written as \cite{berestetskii1980quantum}
\begin{equation}\label{1.11}
    \frac{d \sigma}{d \Omega} = \frac{1}{(4 \pi)^2} \frac{1}{4 M^2} \frac{\vec{k}_2^2}{|\vec{k}_1| \Big( |\vec{k}_2| + \frac{\varepsilon_1}{M} |\vec{k}_2| - \frac{\varepsilon_2}{M} |\vec{k}_1| \cos \theta \Big)} |\widebar{\mathcal{M}}|^2,
\end{equation}
where the energy of the scattered lepton can be reconstructed from the knowledge of the scattering angle and initial parameters as \cite{Gakh201552}
\begin{equation}\label{1.12}
    \varepsilon_2 = \frac {(\varepsilon_1 + M)(\varepsilon_1 M + m^2) + \vec{k}_1^2 \cos \theta \sqrt{M^2 - m^2 \sin^2 \theta}}{(\varepsilon_1 + M)^2 - \vec{k}_1^2 \cos^2 \theta}.
\end{equation}

Finally, one can find the expression for the differential cross section in the Born approximation by plugging Eq. (\ref{1.10}) into Eq. (\ref{1.11}),
\begin{equation}\label{1.13}
\begin{split}
    \frac{d \sigma_{1 \gamma}}{d \Omega} = & \Big[ \frac{G_E^2(Q^2) + \tau G_M^2(Q^2)}{1 + \tau} \\
     + & \frac{Q^2}{4 \varepsilon_1 \varepsilon_2 - Q^2} \Big( 2 \tau - \frac{m^2}{M^2} \Big) G_M^2(Q^2) \Big] \frac{d \sigma_M}{d \Omega} \\
     = & \frac{1}{\epsilon_m (1 + \tau)} \Big[ \tau G_M^2(Q^2) + \epsilon_m G_E^2(Q^2) \Big] \frac{d \sigma_M}{d \Omega},
\end{split}
\end{equation}
where
\begin{equation*}
    4 \varepsilon_1 \varepsilon_2 - Q^2 = \frac{(s-u)^2 - Q^2(4 M^2 +Q^2)}{4 M^2}
\end{equation*}
and the Mott cross section is given by
\begin{equation}\label{1.14}
    \frac{d \sigma_M}{d \Omega} = \frac{\alpha^2}{Q^4} \frac{\Big( 4 \varepsilon_1 \varepsilon_2 - Q^2 \Big) \vec{k}_2^2}{|\vec{k}_1| \Big( |\vec{k}_2| + \frac{\varepsilon_1}{M} |\vec{k}_2| - \frac{\varepsilon_2}{M} |\vec{k}_1| \cos \theta \Big)}
\end{equation}
with the fine-structure constant $\alpha \equiv \frac{e^2}{4 \pi}$ and $Q^2 = 2 (\varepsilon_1 \varepsilon_2 - |\vec{k}_1| |\vec{k}_2| \cos \theta - m^2)$. This is in an agreement with results of Ref. \cite{PhysRevC.36.2466}.

The quantity $\epsilon_m$ describes a measure of the longitudinal polarization of the virtual photon in the UR limit and it can be found to be
\begin{multline}\label{1.15}
    \epsilon_m^{- 1} = 1 - 2 (1 + \tau) \frac {2 m^2 - Q^2}{4 \varepsilon_1 \varepsilon_2 - Q^2} \\
    = \frac{(s-u)^2 + Q^2 (4 M^2 + Q^2) - 4 m^2 (4 M^2 + Q^2)}{(s-u)^2 - Q^2 (4 M^2 + Q^2)}.
\end{multline}
The reason we write  the Born results in the form of Eqs. (\ref{1.13})-(\ref{1.15}) is because they can be easily compared to the well-known UR expressions \cite{PhysRevD.50.5491}
\begin{multline*}
   \Big( \frac{d \sigma_{1 \gamma}}{d \Omega} \Big)_{\mathrm{UR}} = \frac{\Big[ \tau G_M^2(Q^2) + \epsilon G_E^2(Q^2) \Big]}{\epsilon (1 + \tau)} \Big( \frac{d \sigma_M}{d \Omega} \Big)_{\mathrm{UR}},
\end{multline*}
\begin{equation*}
    \Big( \frac{d \sigma_M}{d \Omega} \Big)_{\mathrm{UR}} = \frac{\alpha^2}{4 \varepsilon_1^2} \frac{\cos^2 \frac{\theta}{2}}{\sin^4 \frac{\theta}{2}} \frac{\varepsilon_2}{\varepsilon_1},
\end{equation*}
\begin{equation*}
    \epsilon^{- 1} = \Big( \epsilon_m^{- 1} \Big)_{\mathrm{UR}} = 1 + 2 (1 + \tau) \tan^2 \frac{\theta}{2}.
\end{equation*}

\section{Interference with $\sigma$ exchange}\label{1.100}

The $\sigma$ meson is the lightest scalar meson observed in nature, and it describes a medium-range nucleon-nucleon attraction  \cite{PhysRevC.63.024001, PhysRevC.51.38}, which is responsible for the nuclear binding. The precise position of $\sigma$'s pole is difficult to establish because it has a large decay width and because it cannot be explained by a naive Breit-Wigner resonance. For these reasons, a considerable number of different models exists to characterize the properties of $\sigma$. These models give us various estimations of the coupling of $\sigma$ to the proton. In our calculations, we choose to follow the predictions of models \cite{ANDP:ANDP201400077} and \cite{Downum2006455}. The former approach uses the effective hadron Lagrangian in the quark model to calculate $\sigma N N$ coupling. The latter one, in its turn, employs the knowledge about the Compton scattering amplitudes and the electromagnetic polarizabilities. As a result, these models provide us with following couplings to the proton: $g_{\sigma p p} = 3 \div 7$ ($m_\sigma = 500$ MeV) and $g_{\sigma p p} = 13.1 \div 13.2$ ($m_\sigma = 666$ MeV), respectively.

The second term in, Eq. (\ref{1.7}), describes the interference between the left and right diagrams of Fig.\ref{fig:1}. One can show (see Appendix \ref{1.300}) that the appropriate matrix element describing $l^\pm$ scattering off of the proton is given by
\begin{multline}\label{1.16}
    2 \ \mathrm{Re} [\widebar{\mathcal{M}^{v}{\mathcal{M}^s}^*}] = \\
    = \mp \frac {8 m M (s - u) e^2}{Q^2 (Q^2 + m^2_\sigma)} G_E(Q^2) g_{\sigma p p} \ \mathrm{Re}[f_s].
\end{multline}

Note that  the obtained expression, Eq. (\ref{1.16}), is proportional to the real part of $\sigma$'s coupling to the lepton as well as to the mass of the lepton, which means that it is going to be enhanced considerably for the muons.

\begin{figure}
	\includegraphics[scale=0.45]{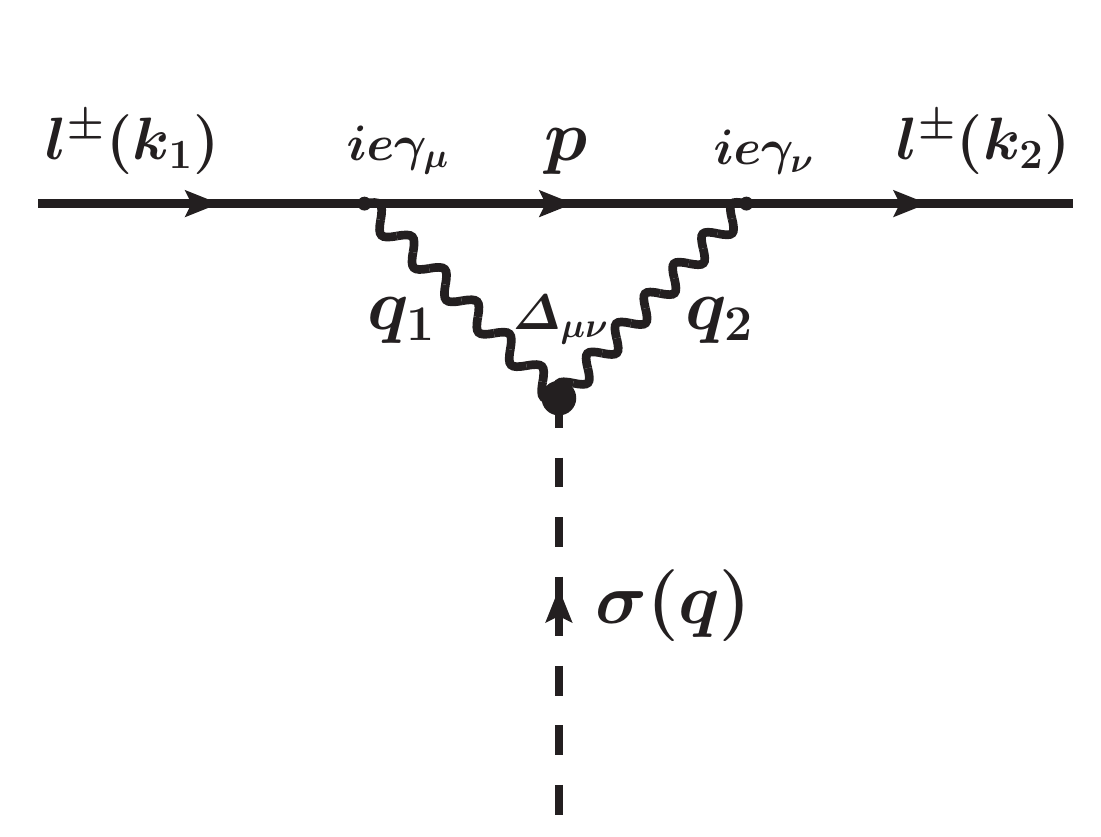}
	\caption{Coupling of $\sigma$ meson to the lepton via two-photon conversion.}
    \label{fig:2}
\end{figure}

The form factor $f_s$ can be found by considering all the possible ways in which $\sigma$ can couple to the lepton. We claim that the dominant contribution there will be provided by the diagram shown in Fig. \ref{fig:2}. Having the coupling of $\sigma$ to the lepton identified, we can construct the corresponding amplitude
\begin{equation}\label{1.18}
\begin{split}
    T = & i e^4 \int \frac {d^4 p}{(2 \pi)^4} \bar{u}(k_2) \frac {\gamma_\nu (\slashed{p} + m) \gamma_\mu}{p^2 - m^2} u(k_1) \frac{1}{q^2_1} \Delta_{\mu \nu} \frac{1}{q^2_2} \\
    \equiv & \bar{u}(k_2) f_s u(k_1),
\end{split}
\end{equation}
where the integration is performed over the momentum $p$ of the intermediate lepton, $q_1$ and $q_2$ are the momenta of exchange photons that carry polarizations $\mu$ and $\nu$, respectively. $\Delta_{\mu\nu}$ depicts the coupling of virtual $\sigma$ to two virtual photons. The most general form for $\Delta_{\mu\nu}$ is given in \cite{Dorokhov2012}
\begin{multline}\label{1.200}
    \Delta_{\mu \nu} = A(q^2, q^2_1, q^2_2) \Big( g_{\mu \nu} (q_1 \cdot q_2) - q_{1 \nu} q_{2 \mu} \Big) + B(q^2, q^2_1, q^2_2) \\
    \times \Big( q^2_1 q_{2 \mu} - (q_1 \cdot q_2) q_{1 \mu} \Big) \Big( q^2_2 q_{1 \nu} - (q_1 \cdot q_2) q_{2 \nu} \Big).
\end{multline}

As we can see, this coupling consists of two terms: the first term represents transverse photons exchanges, and the second term represents longitudinal photons exchanges. The latter one includes the form factor $B(q^2, q^2_1, q^2_2)$, which is challenging to model due to experimental inability to measure longitudinal photons' contribution. Since the main goal of this article is to estimate the leading helicity-flip contribution for the scattering of muons in the given kinematics (related to future MUSE measurements), we will consider only the contribution from transverse photons and neglect the contribution from longitudinal photons (that may be of the same order).

To calculate the first term in Eq. (\ref{1.200}) we use the vector meson dominance (VMD) model to depict the corresponding form factor $A(q^2, q^2_1, q^2_2)$. Here, we should note that we assume that the $\sigma$ meson couples each photon via a $\rho$ meson. As a result, the $\Delta_{\mu\nu}$ vertex takes the form
\begin{equation}\label{1.19}
    \Delta_{\mu\nu} = \frac{g_{\sigma \gamma \gamma}}{\Big( 1 - \frac {q^2_1}{m^2_\rho} \Big) \Big( 1 - \frac {q^2_2}{m^2_\rho} \Big)} \Big[ g_{\mu\nu} (q_1 \cdot q_2) - q_{1\nu} q_{2\mu} \Big],
\end{equation}
where $m_\rho$ is the mass of the vector $\rho$ meson and $g_{\sigma \gamma \gamma}$ is the $\sigma$ meson coupling constant to two real photons. $g_{\sigma \gamma \gamma}$ can be found from the knowledge of the decay width $\Gamma_{\sigma \rightarrow \gamma \gamma}$. Both quantities are related via
\begin{equation}\label{1.20}
    g_{\sigma \gamma \gamma} = \Big( \frac {4 \ \Gamma_{\sigma \rightarrow \gamma \gamma}}{\pi \alpha^2 m^3_\sigma} \Big)^{1/2}.
\end{equation}

We do not fix a value of the decay width $\Gamma_{\sigma \rightarrow \gamma \gamma}$, but instead use a range obtained in the partial wave amplitudes analysis \cite{Dai201411} $\Gamma_{\sigma \rightarrow \gamma \gamma} = 1.8 \div 2.3$ keV ($m_\sigma = 500$ MeV) as well as the range obtained in the Compton scattering analysis \cite{ANDP:ANDP201400077} $\Gamma_{\sigma \rightarrow \gamma \gamma} = 2.3 \div 2.9$ keV ($m_\sigma = 666$ MeV).

\begin{figure}
	\includegraphics[scale=0.65]{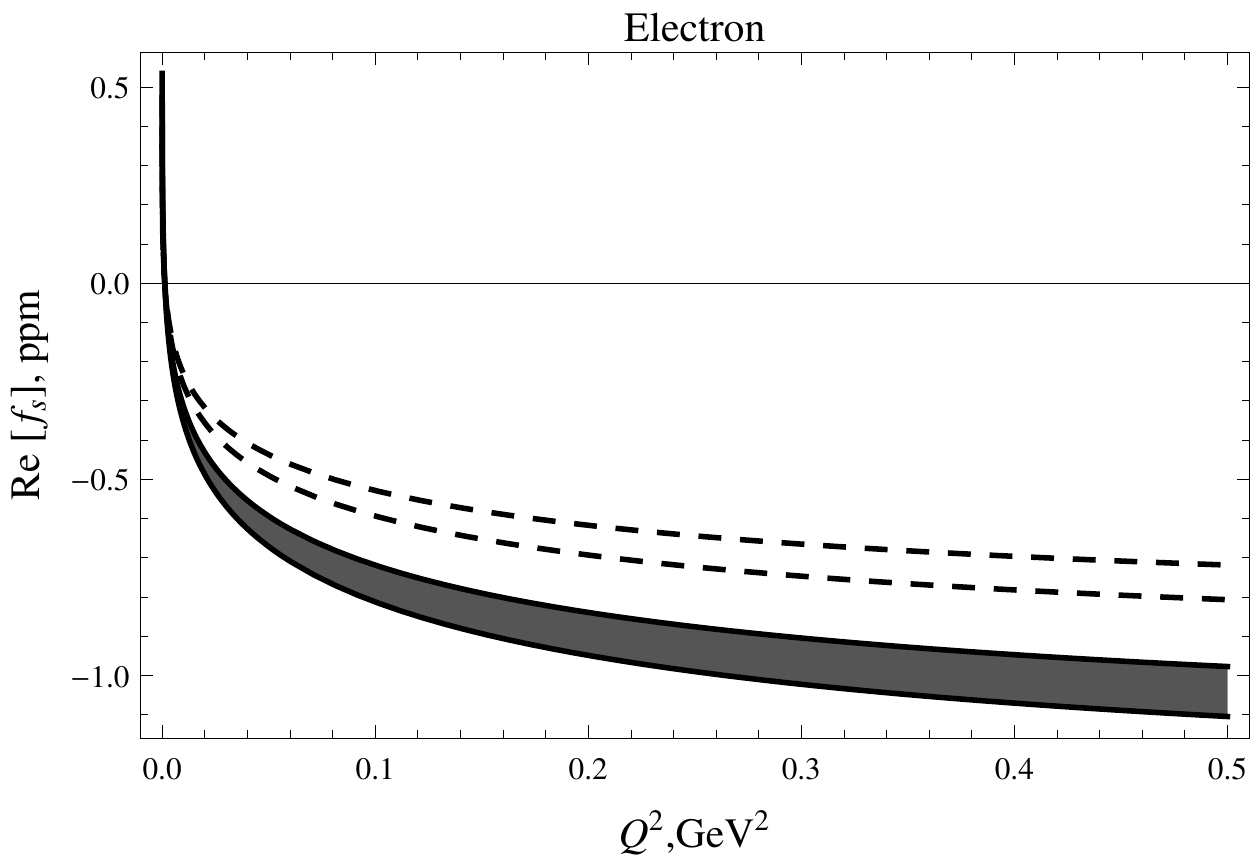}
	\caption{Electron coupling form factor $f_s$ for $\Gamma_{\sigma \rightarrow \gamma \gamma} = 1.8 \div 2.3$ keV  \cite{Dai201411} (shaded region, solid lines, $m_\sigma = 500$ MeV) and $\Gamma_{\sigma \rightarrow \gamma \gamma} = 2.3 \div 2.9$ keV \cite{ANDP:ANDP201400077} (transparent region, dashed lines, $m_\sigma = 666$ MeV)}
    \label{fig:3}
\end{figure}

\begin{figure}
	\includegraphics[scale=0.64]{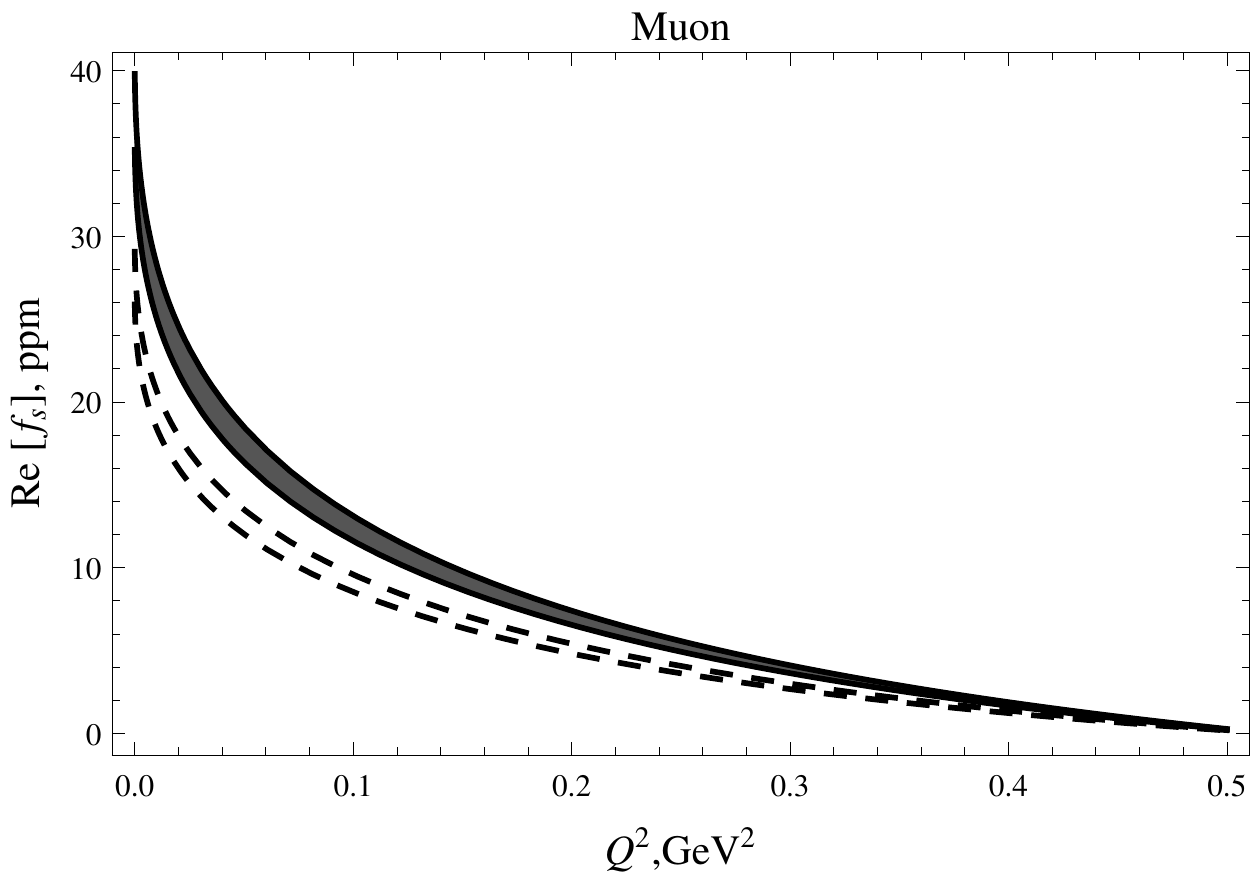}
	\caption{Muon coupling form factor $f_s$ for $\Gamma_{\sigma \rightarrow \gamma \gamma} = 1.8 \div 2.3$ keV  \cite{Dai201411} (shaded region, solid lines, $m_\sigma = 500$ MeV), and $\Gamma_{\sigma \rightarrow \gamma \gamma} = 2.3 \div 2.9$ keV \cite{ANDP:ANDP201400077} (transparent region, dashed lines, $m_\sigma = 666$ MeV)}
    \label{fig:4}
\end{figure}

In VMD, the amplitude $T$ in Eq. (\ref{1.18}) takes the form of Passarino-Veltman's five-point functions. Using the underlying identity from Ref. \cite{PhysRevD.55.4380} and applying it twice, as necessary,
\begin{equation*}
    \frac{1}{A B} = \frac{1}{B-A} \Big[ \frac{1}{A} - \frac{1}{B} \Big]
\end{equation*}
one can show that the five-point function in Eq. (\ref{1.18}) can be reduced to the sum of standard three-point functions due to the fact that only two of 4-momenta in the $\sigma\gamma\gamma$ vertex are independent. We calculate these three-point functions numerically using the LoopTools software \cite{HAHN1999153} and obtain the form factor's $f_s$ dependence on $Q^2$  shown in Figs. \ref{fig:3} and \ref{fig:4}.

It should be noted that the $\Delta_{\mu\nu}$ vertex in VMD falls off as $\sim Q^{-4}$ at high momentum transfers. This is consistent with the asymptotic scaling rules \cite{PhysRevD.28.228} if the $\sigma$ meson is viewed as a quasibound state of two pions (or a $qq\bar{q}\bar{q}$ state).

Once the form factor $f_s$ was evaluated, the interference cross section $\frac{d \sigma_I}{d \Omega}$ was found by plugging Eq. (\ref{1.16}) into Eq. (\ref{1.11}). The interference contribution is charge dependent. Therefore, to find the difference between elastic $l^+ - p$ and $l^- - p$ scattering it is convenient to define
\begin{equation}\label{1.21}
    \frac{d \sigma^\pm}{d \Omega} = \frac{d \sigma_{1 \gamma}}{d \Omega} \pm \frac{d \sigma_I}{d \Omega} \equiv \frac{d \sigma_{1 \gamma}}{d \Omega} (1 \pm \delta).
\end{equation}

Then, the asymmetry $A$ is given exactly by $\delta$,
\begin{equation}\label{1.22}
    A \equiv \frac {\frac{d \sigma^+}{d \Omega} - \frac{d \sigma^-}{d \Omega}}{\frac{d \sigma^+}{d \Omega} + \frac{d \sigma^-}{d \Omega}} = \delta.
\end{equation}

\begin{figure*}[htp]
	\includegraphics[scale=1]{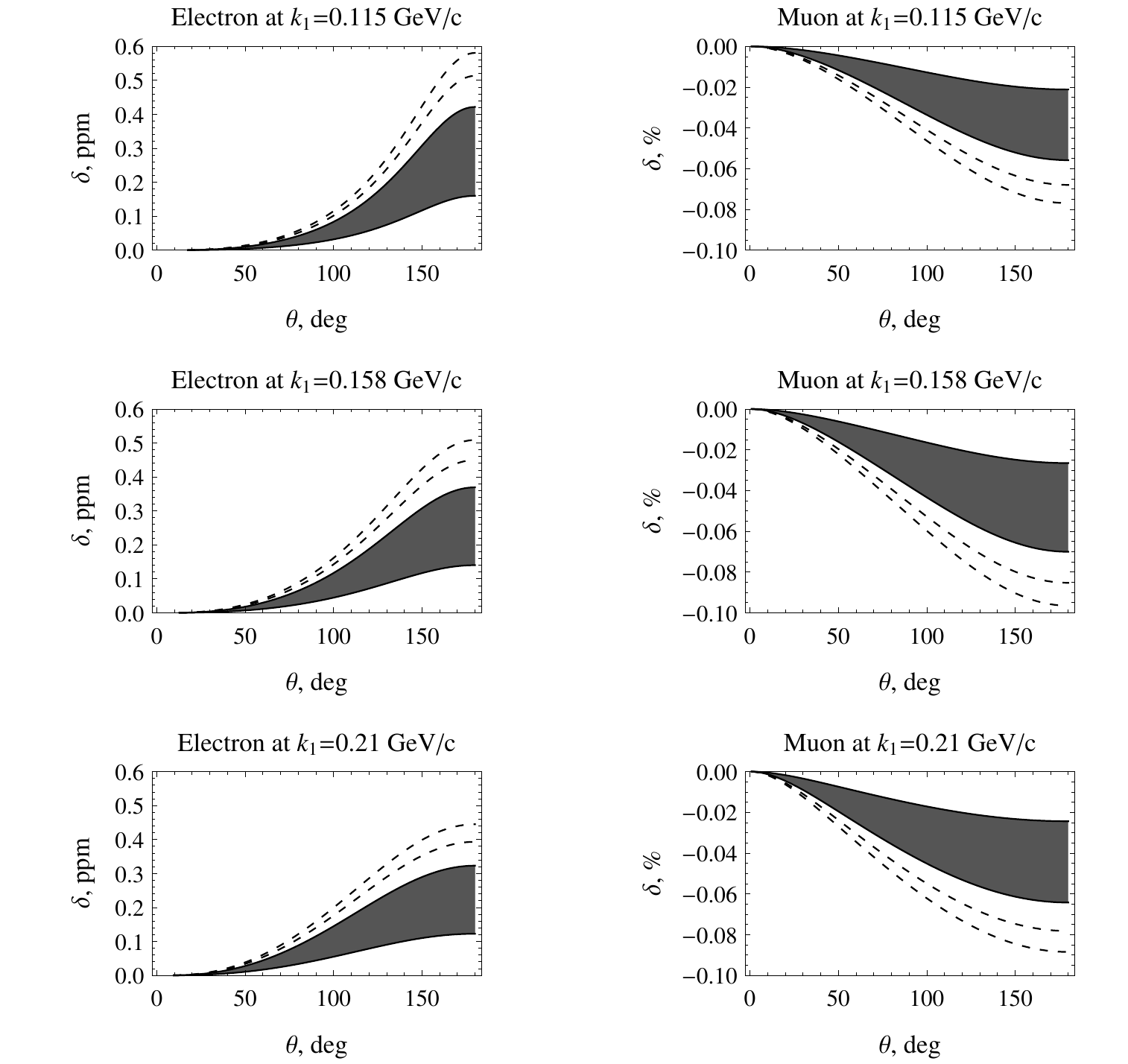}
	\caption{Asymmetry for $\Gamma_{\sigma \rightarrow \gamma \gamma} = 1.8 \div 2.3$ keV, $g_{\sigma p p} = 3 \div 7$ (shaded region, solid lines, $m_\sigma = 500$ MeV), and $\Gamma_{\sigma \rightarrow \gamma \gamma} = 2.3 \div 2.9$ keV, $g_{\sigma p p} = 13.1 \div 13.2$ (transparent region, dashed lines, $m_\sigma = 666$ MeV)}
    \label{fig:5}
\end{figure*}

The corresponding angular dependence of $\delta$ is shown in Fig. \ref{fig:5} at the momenta of MUSE. It is worth mentioning that the value of $g_{\sigma p p}$ coupling, in general, depends on the momentum transfer squared. However, this dependence affects negligibly the final calculations in our $Q^2$ range. For instance, if we choose the well-tested OBE potential dependence \cite{PhysRevC.63.024001} as the model, our final result changes relatively only by $0.4\%$.

Analytically, one can find that
\begin{multline}\label{1.23}
    \delta = - \frac{m}{\pi \alpha} \ g_{\sigma p p} \ G_E(Q^2) \ \mathrm{Re}[f_s] \\
    \times \frac{\sqrt{\tau (1 + \tau) (1 - \tilde{\epsilon}^2_m) M^2 + m^2 (1 + \tau)(1 - \tilde{\epsilon}_m)^2 }}{[ (\tilde{\epsilon}_m + \eta) G^2_E (Q^2) + \tau G^2_M (Q^2)] (Q^2 + m^2_\sigma)} ,
\end{multline}
where
\begin{equation}
    \eta = \frac{2 m^2}{Q^2} (1 - \tilde{\epsilon}_m),
\end{equation}
\begin{equation}
\begin{split}
    \tilde{\epsilon}_m = & \frac{(s-u)^2 - Q^2 (4 M^2 + Q^2) - 4 m^2 (4 M^2 + Q^2)}{(s-u)^2 + Q^2 (4 M^2 + Q^2) - 4 m^2 (4 M^2 + Q^2)} \\
    = & \frac{Q^2 \epsilon_m - 2 m^2}{Q^2 - 2 m^2}.
\end{split}
\end{equation}
Note that $\tilde{\epsilon}_m = 1$ at $Q^2 = 0$ and $\tilde{\epsilon}_m \rightarrow 0$ at $Q^2 = Q^2_{max}$.

\section{Conclusions and Discussions}
In this paper, we calculated the contribution from the scalar $\sigma$ meson exchange to the differential cross section of elastic lepton-proton scattering. To obtain the result, we revised the ultrarelativistic Born approximation calculations, taking into account a nonzero mass of the lepton. Our main finding is that the $\sigma$ exchange contribution is about 3 orders in magnitude larger for the massive muon than for the much lighter electron in the kinematics of MUSE. Our result can be treated as an additional and independent contribution to TPE calculations of Ref. \cite{PhysRevD.90.013006}, which were performed under the assumption of the elastic (proton) intermediate state. Our contribution appears to be on the order of $\lesssim 0.1 \%$ for muons. It is comparable in magnitude with the inelastic contribution that was calculated in Ref. \cite{Tomalak2016} by employing the forward doubly virtual Compton scattering approximation, and it is about five times smaller than the leading (elastic) contribution.

In summary, we have evaluated explicitly the contribution of $t$-channel $\sigma$ meson exchange to TPE. This contribution is proportional to the lepton mass, and therefore it is strongly enhanced for muons and suppressed for the electrons in MUSE kinematics.

The obtained result was compared with the unpolarized lepton-proton scattering predictions of Ref. \cite{Liu.Miller.2015}. In that paper, the authors discuss the effects due to a light scalar boson exchange with $m_\phi \sim 1$ MeV. In particular, they have found that for the scattering of muons the corresponding interference contribution is on the order of $10^{-6}$. The boson considered there possesses exactly the same properties as the $\sigma$ meson, besides the fact that the mass of the boson and the mass of the $\sigma$ meson are different. Therefore, it was not difficult for us to check our calculations for $m_\sigma = 1$ MeV. Our estimations are in a good agreement with Ref. \cite{Liu.Miller.2015}. It should be noted that the increased contribution due to the smaller mass in the bosons' propagator is compensated by the smaller value of the coupling $g_{\phi p p}$.

The main impediment to performing our calculations was the lack of knowledge about the virtual $\sigma$ meson coupling to two photons. In this work, we performed an estimation of this coupling accounting only for the coupling to the transverse photons in the vector meson dominance model. The prediction showed little sensitivity to the momentum dependence of $g_{\sigma p p}$ coupling in the considered kinematics.

\appendix
\section{}\label{1.300}

In this Appendix we provide the derivation of the interference between vector ($\gamma$-exchange) and scalar ($\sigma$ meson exchange) currents as well as between vector and pseudoscalar ($\pi$ meson exchange) currents for the unpolarized lepton scattering off of the proton target. Both contributions can be calculated by using expressions for the currents given in Eqs. (\ref{1.3}) - (\ref{1.5}) and by following the standard procedure of summing over the final and averaging over initial spin states of the particles. As a result, one can find that

\begin{widetext}
\begin{equation}\label{1.25}
\begin{split}
2 \ \mathrm{Re} [\widebar{j^{v}_{\mu} J^{v}_{\mu} (j^s J^s)^*}] = \frac {1}{4} f^*_s \ g^*_s \Big( & F_1(Q^2) \ \mathrm{Tr} \big[(\slashed{k}_2 + m) \gamma_\mu (\slashed{k}_1 + m) \big] \ \mathrm{Tr} \big[(\slashed{p}_2 + M) \gamma_\mu (\slashed{p}_1 + M) \big] \\
+ & \frac {i}{2 M} F_2 (Q^2) \ \mathrm{Tr} \big[(\slashed{k}_2 + m) \gamma_\mu (\slashed{k}_1 + m) \big] \ \mathrm{Tr} \big[(\slashed{p}_2 + M) \sigma_{\mu \nu} q_\nu (\slashed{p}_1 + M) \big] \Big) \\
+ \frac {1}{4} f_s \ g_s \Big( & F^*_1(Q^2) \ \mathrm{Tr} \big[(\slashed{k}_2 + m) (\slashed{k}_1 + m) \gamma_\mu \big] \ \mathrm{Tr} \big[(\slashed{p}_2 + M) (\slashed{p}_1 + M) \gamma_\mu \big] \\
+ & \frac {i}{2 M} F^*_2 (Q^2) \ \mathrm{Tr} \big[(\slashed{k}_2 + m) (\slashed{k}_1 + m) \gamma_\mu \big] \ \mathrm{Tr} \big[(\slashed{p}_2 + M) (\slashed{p}_1 + M) \sigma_{\mu \nu} q_\nu \big] \Big),
\end{split}
\end{equation}
\end{widetext}

\begin{widetext}
\begin{equation}\label{1.26}
\begin{split}
2 \ \mathrm{Re} [\widebar{j^{v}_{\mu} J^{v}_{\mu} (j^p J^p)^*}] = \frac {1}{4} f^*_p \ g^*_p \Big( & F_1(Q^2) \ \mathrm{Tr} \big[(\slashed{k}_2 + m) \gamma_\mu (\slashed{k}_1 + m) \gamma_5 \big] \ \mathrm{Tr} \big[(\slashed{p}_2 + M) \gamma_\mu (\slashed{p}_1 + M) \gamma_5 \big] \\
+ & \frac {i}{2 M} F_2 (Q^2) \ \mathrm{Tr} \big[(\slashed{k}_2 + m) \gamma_\mu (\slashed{k}_1 + m) \gamma_5 \big] \ \mathrm{Tr} \big[(\slashed{p}_2 + M) \sigma_{\mu \nu} q_\nu (\slashed{p}_1 + M) \gamma_5 \big] \Big) \\
+ \frac {1}{4} f_p \ g_p \Big( & F^*_1(Q^2) \ \mathrm{Tr} \big[(\slashed{k}_2 + m) \gamma_5 (\slashed{k}_1 + m) \gamma_\mu \big] \ \mathrm{Tr} \big[(\slashed{p}_2 + M) \gamma_5 (\slashed{p}_1 + M) \gamma_\mu \big] \\
+ & \frac {i}{2 M} F^*_2 (Q^2) \ \mathrm{Tr} \big[(\slashed{k}_2 + m) \gamma_5 (\slashed{k}_1 + m) \gamma_\mu \big] \ \mathrm{Tr} \big[(\slashed{p}_2 + M) \gamma_5 (\slashed{p}_1 + M) \sigma_{\mu \nu} q_\nu \big] \Big).
\end{split}
\end{equation}
\end{widetext}

\begin{widetext}
By calculating the traces above, employing the fact that Dirac and Pauli form factors are real functions of $Q^2$, and using the definition Eq. (\ref{1.110}) one can show that
\end{widetext}

\begin{widetext}
\begin{equation}\label{1.27}
\begin{split}
2 \ \mathrm{Re} [\widebar{j^{v}_{\mu} J^{v}_{\mu} (j^s J^s)^*}] = & 8 g_s  \Big[ m M \Big( 2 s - Q^2 - 2 m^2 - 2 M^2 \Big) F_1(Q^2) - \frac{m}{2 M} Q^2 \Big( s - m^2 - M^2 - \frac{1}{2} Q^2 \Big) F_2(Q^2) \Big] \ \mathrm{Re}[f_s] \\
= & 8 m M (s - u) \ G_E(Q^2) g_s \mathrm{Re}[f_s], \\
2 \ \mathrm{Re} [\widebar{j^{v}_{\mu} J^{v}_{\mu} (j^p J^p)^*}] = & 0.
\end{split}
\end{equation}
\end{widetext}

\begin{acknowledgments}
We thank W. J. Briscoe, E. J. Downie, H. W. Griesshammer, and M. R. Pennington for useful discussions. This work was supported by the NSF under Grants No. PHY-1404342 and No. PHY-1309130 and by The George Washington University through the Gus Weiss endowment.
\end{acknowledgments}

\bibliography{article_sigma}

\end{document}